\documentclass[fleqn,twoside]{article}
\usepackage{espcrc2}

\usepackage{graphicx}
\usepackage[figuresright]{rotating}

\newcommand{\AmS}{{\protect\the\textfont2
  A\kern-.1667em\lower.5ex\hbox{M}\kern-.125emS}}

\title{Reactor measurement of $\theta_{12}$; Secret of the power
}

\author{H. Minakata\address{Department of Physics, 
Tokyo Metropolitan University, Hachioji, Tokyo 192-0397, Japan}\thanks{
Based on presentation at Neutrino Oscillation Workshop (NOW2004), 
September 11-17, 2004, Conca Specchiulla, Otranto, Italy}
H. Nunokawa\address{Pontif{\'\i}cia Universidade Cat{\'o}lica 
do Rio de Janeiro, C. P. 38071, 22452-970, Rio de Janeiro, Brazil}
W. J. C. Teves\address[USP]{Instituto de F{\'\i}sica,  Universidade de S{\~a}o Paulo, 
C.\ P.\ 66.318, 05315-970, S{\~a}o Paulo, Brazil}
and
R. Zukanovich Funchal\addressmark[USP]
}

\begin{document}

\begin{abstract}

We demonstrate enormous power of dedicated reactor experiment 
for $\theta_{12}$ with a detector placed at around the first oscillation 
maximum, which we call ``SADO". 
It allows determination of $\sin^2\theta_{12}$ to the accuracy of 
$\simeq$2\% at 1$\sigma$ CL, 
which surpasses all the method so far proposed. 
Unlike reactor $\theta_{13}$ experiments, the requirement for the 
systematic error is very mild, $\simeq$4\%, which makes it an even more 
feasible experiment. 
If we place a detector at $\sim$60 km away from the Kashiwazaki-Kariwa 
nuclear power plant, 0.5 kt$\cdot$yr exposure of SADO is equivalent 
to $\sim$100 kt$\cdot$yr exposure of KamLAND assuming the same 
systematic error.

%\vspace{1pc}
\end{abstract}

\maketitle

\section{Introduction}
\label{sec:intro}

After the three neutrino experiments, SK  \cite{SKatm_new}, 
K2K \cite{K2K_new}, and KamLAND \cite{KamLAND_new},  
saw the oscillatory behavior with atmospheric and solar neutrino 
mixing parameters, physics of neutrino oscillation has entered 
into a new era. 
NOW2004 is so timely held that it is the first dedicated 
workshop to neutrino oscillations in the new era.

One of the directions which will be pursued in the new era 
is the precise determination of the lepton mixing parameters.
Suppose that in someday we succeed to construct the 
``standard model of flavor mixing".
Given the enormous development of particle physics in the 
last 30 years, it is highly unlikely that it will never happen. 
Less ambitious assumption is that there might be some relationship 
between the quark and the lepton mixing parameters, whose example is 
given by the quark-lepton complementarity \cite{QLC}.
In trying to test such theories, then, we will discover the great 
disparity between accuracies of measurement of mixing parameters 
in the quark and the lepton sectors.

The most accurately measured quark mixing angle 
is the Cabibbo angle, whose error is about 1.4\% in $\sin^2\theta_{C}$ 
at 90\% CL \cite{PDG}. 
The most accurately measured lepton mixing angle is the 
solar angle, $\theta_{12}$, whose error is about 14\% in 
$\sin^2\theta_{12}$ at the same CL \cite{global}. 
Future operation of the solar and the KamLAND experiments 
are expected to lead to improvement only by a factor of $\sim$2 
\cite{bahcall-pena}. 
So the right question is: 
``are there ways to measure $\theta_{12}$ far more accurately than it is now?" and 
``can the sensitivity ever reach to the level comparable to the Cabibbo angle?"
We answer in the positive to these questions.

It is well known for years that the best way to achieve 
optimal sensitivity for mixing angles is to exploit 
energy and baseline tuned to the oscillation maximum. 
(For a pedagogical exposition of this principle, see \cite{MNTZ}.)
For small mixing angle $\theta_{13}$, it gives the highest chance 
of accessing to the unique unknown mixing angle, and the method 
has been widely exploited by the reactor $\theta_{13}$ experiments 
\cite{MSYIS,reactor_white}. 
For large mixing angles $\theta_{12}$ and $\theta_{23}$, 
on the other hand, it allows us the best hope for realizing 
the highest sensitivity. 
In the JPARC-SK experiment, the principle is utilized to reach to the 
sensitivity as high as $\simeq$1-2\% in $\sin^22\theta_{23}$ 
\cite{JPARC,MSS}.\footnote{
%%%%%%%%%%%%%%%  footnote %%%%%%%%%%%%%%%%%
In the talk at the workshop the problem of accurate determination of 
$\theta_{23}$ was also addressed. But, we concentrate on 
$\theta_{12}$ in this manuscript because the space is quite limited. 
We refer \cite{MSS} for the latter topics.
}

Then, it is entirely natural to try to extend the method of tuning to 
the oscillation maximum to precision $\theta_{12}$ measurement, 
and it is what we discuss in this manuscript based on \cite{MNTZ}. 
See \cite{BCGP} for a similar but different proposal.
For our purpose, the right distance is of order 
$L=L_{\rm OM} \equiv 2\pi E_{\rm peak}/\Delta m^2_{21} \simeq 60$ km, 
where $E_{\rm peak}=4$ MeV is a peak energy of event spectrum.  
Let us call, for ease of frequent reference, a detector placed at 
around the oscillation maximum ``SADO'', an acronym of 
Several-tens of km Antineutrino DetectOr. 
Though our discussion is fully based on the results obtained in 
\cite{MNTZ}, we will present informations complementary to it. 
In particular, all the figures are new.

\section{Requirements for the experimental systematic error}

In order to design feasible experiments we cannot be too 
optimistic to the experimental systematic error. 
Because energy spectrum cut at $E_{\rm prompt}=2.6$ MeV 
produces the systematic error of 2.3\% in KamLAND \cite{KamLAND_new}, 
it is better not to do spectrum cut. But, then, we have to deal with 
geo-neutrino contamination, the problem addressed in detail in 
\cite{MNTZ}. 
Fortunately, we have uncovered that we can accommodate a rather 
relaxed value 4\% of the experimental systematic error.
It should be within reach, given the current KamLAND error of 6.5\%, 
if the fiducial volume error is better controlled and no spectrum 
cut is performed.
It is also found that geo-neutrino contamination can be tolerable 
by an appropriate choice of the baseline, $L=50-70$ km.

\section{SADO sensitivity of $\theta_{12}$}

\begin{figure}[htb]
\begin{center}
\includegraphics[width=18pc]{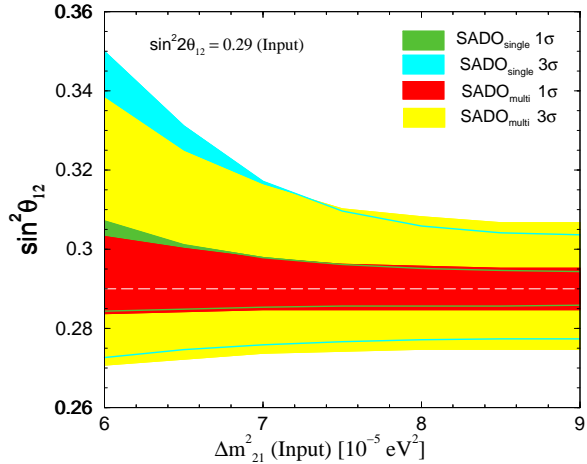}
\vspace{-1.0cm}
\caption{The error of $\sin^2\theta_{12}$ (1 degree of freedom (DOF)) 
as a function of $\Delta m^2_{21}$ 
expected by SADO at 54 km away from Kashiwazaki-Kariwa nuclear 
power plant with 60 GW$_{th}\cdot$kt$\cdot$yr exposure.
SADO$_{\rm multi}$ and SADO$_{\rm single}$ refer, respectively, the cases 
with and without other 15 reactors.
The geo-neutrinos are treated by the Fully Radiogenic model  \cite{MNTZ}.
}
\vspace{-0.7cm}
\label{dm2dep}
\end{center}
\end{figure}

Now we present SADO sensitivity of $\theta_{12}$ in Fig.~\ref{dm2dep}. 
To make it complementary to the one given in \cite{MNTZ}, 
we plot the errors of $\sin^2\theta_{12}$ as a function of 
$\Delta m^2_{21}$. 
Because of possible change in the best fit value of $\Delta m^2_{21}$ 
in the future, it would be useful to have such an information. 
Since the vacuum oscillation probability is a function of $\Delta m^2_{21} L$, 
the information is (not quite equivalent but) 
complementary to the 
$L$-dependence of the error given in Fig.3 of  \cite{MNTZ}.

%%%%%%%%%%%%%%%% Fig.1 %%%%%%%%%%%%%%%%%

As we see in Fig.~\ref{dm2dep}, the SADO sensitivity can reach to 
$\simeq$2\% in $\sin^2\theta_{12}$ at 1 $\sigma$ CL 
($\simeq$3\% in 90\% CL).
Therefore, it can reach to the level roughly 
comparable to that of the Cabibbo angle. 
It is also remarkable that the sensitivity remains the same within 
$\pm$20\% for a wide range of $\Delta m^2_{21}$
currently allowed by the KamLAND data. 
It in turn implies that a wide range of baseline,  
50-70 km from the reactor complex, is suitable, 
allowing a variety of possibilities for site selection.

For remaining issues like 
dependence of sensitivity on geo-neutrino models and running time, 
the effects of surrounding reactors,  
as well as detailed analysis procedure with a careful stability check 
of the statistical method,  see \cite{MNTZ}.

\section{KamLAND vs. SADO}

%%%%%%%%%%%%%%%% Fig.2 %%%%%%%%%%%%%%%%%
\begin{figure}[htb]
\begin{center}
\includegraphics[width=18pc]{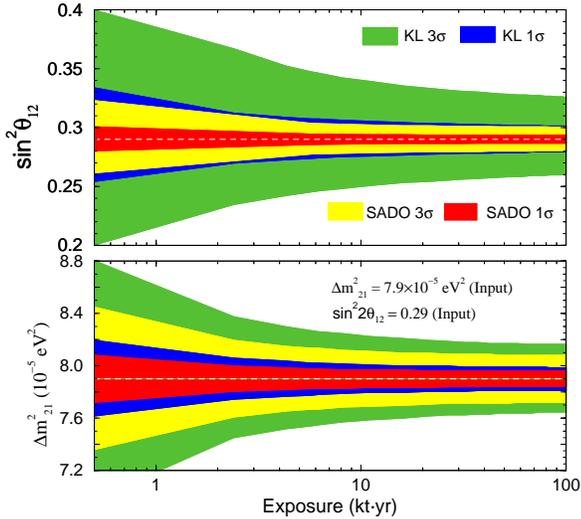}
\vspace{-1.0cm}
\caption{Accuracies of determination of 
$\sin^2\theta_{12}$ (upper panel) and 
$\Delta m^2_{21}$ (lower panel) reachable by 
KamLAND and SADO (both 1 DOF) are compared with the 
same systematic error of 4\%. 
The geo-neutrino contribution was switched off.
}
\vspace{-0.8cm}
\label{dist-dep}
\end{center}
\end{figure}

KamLAND is the marvelous experiment that has settled the 
solar neutrino problem which lasted for nearly 40 years by 
observing deficit of antineutrinos from reactors located at 
100-200 km from Kamioka \cite{KamLAND}. 
It will run $\sim$10 more years in the future.
Therefore, unless SADO supersedes the sensitivity of KamLAND 
with large margin there is no sense of talking about such an 
expensive new project. 
Moreover, there is a real merit of the KamLAND-SADO comparison;  
It should prove (or disprove) how powerful is the method of 
tuning the baseline distance.

We present in Fig.~\ref{dist-dep} KamLAND vs. SADO comparison 
of the sensitivities as a function of kt$\cdot$yr, assuming that 
SADO is placed at 54 km away from the Kashiwazaki-Kariwa 
nuclear power plant. 
We observe that to reach the same accuracy of $\sin^2\theta_{12}$ 
achieved by SADO in a short exposure of 0.5 kt$\cdot$yr, KamLAND 
would take $\sim$100 kt$\cdot$yr.
Notice that both detectors receive the same neutrino flux from 
all the 16 reactors in Japan. Therefore, the difference in their 
sensitivities reflects just their locations and the result testifies 
for the power of the method of tuning to the first oscillation maximum.

\section{Solar plus KamLAND vs. SADO}

%%%%%%%%%%%%%%%% Fig.3 %%%%%%%%%%%%%%%%%
\begin{figure}[htb]
\vspace{-2.2cm}
\begin{center}
\includegraphics[width=22pc]{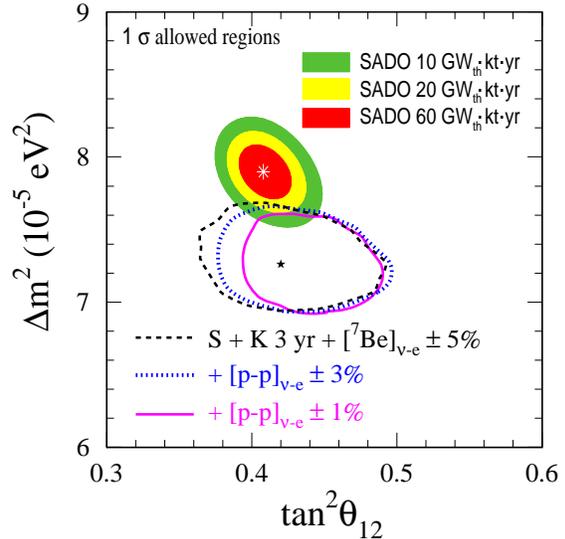}
\vspace{-5.0cm}
\caption{SADO's sensitivity contours are plotted in 
$\tan^2\theta_{12}$-$\Delta m^2_{21}$ space and are 
overlaid on Fig.6 of \cite{bahcall-pena}, in which the sensitivities 
of solar-KamLAND combined method are presented. 
The errors are defined both with 2 DOF.
}
\vspace{-0.7cm}
\label{solarKL-SADO}
\end{center}
\end{figure}

A burning question by the readers may be whether 
SADO can surpasses the great sensitivity to be achieved jointly 
by KamLAND and precision measurement of solar 
neutrino fluxes, in particular, $^7$Be and pp neutrinos. 
In Fig.~\ref{solarKL-SADO}, we present the results 
obtained by Bahcall and Pe{\~n}a-Garay \cite{bahcall-pena}, and 
the corresponding sensitivity to be achieved by SADO \cite{MNTZ}, 
both plotted in terms of $\tan^2\theta_{12}$ following \cite{bahcall-pena}.
As you can observe from the figure, SADO has potential of 
superseding the solar-KamLAND combined method in sensitivities 
not only to $\theta_{12}$ but also to $\Delta m^2_{21}$.
Notice that the solid (purple) line of the solar-KamLAND method 
assumes total experimental error of 1\% for $pp$ neutrino 
observation, and therefore it may be called as an ultimate 
accuracy achievable by the method.

It must be emphasized that the better accuracy of 
$\theta_{12}$ by SADO {\it does not} lower the value of 
planned low-energy solar neutrino experiments. 
See \cite{lowE} for overview of such experiments.
Since the uncertainty of $\theta_{12}$ may be the largest 
source of the systematic error in such experiments, SADO 
does indeed helps them by decreasing the major part of the 
systematic error. Therefore, it will facilitate highly accurate 
determination of solar neutrino fluxes, in particular the $pp$ flux, 
which is most important to probe structure of the principal solar engine.

\section{Physics implications}

Precision measurement of $\theta_{12}$ has a number of 
interesting physics implications not only in particle physics point of view 
but also in observational solar astrophysics and geo-physics. 
It will open a new era of lepton mixing parameter determination 
with the accuracy comparable with quark sector.
In observational solar physics, precision measurement of 
$\theta_{12}$ plays a key role for accurate flux determination, 
and is indispensable for model-independent test of the 
standard solar model.
In geo-physics context, it will help to remove "reactor background" 
of geo-neutrino measurement at KamLAND and at Borexino \cite{borexino}. 
It will contribute to future neutrino experiments, e.g., by 
decreasing the ambiguity in observing CP violation effect 
which comes from uncertainties of other mixing parameters.  
It improves the accuracy of the test of CPT invariance in the lepton sector 
and various exotic hypotheses including flavor changing neutral current. 
See \cite{MNTZ}  for more details.

\section{Concluding remarks}

In the talk we emphasized that there exists a coherent view for 
measuring the large mixing angles, $\theta_{12}$ and 
$\theta_{23}$, precisely by tuning the baseline distance.  
Quite unfortunately, in the case of $\theta_{23}$ the enormous 
accuracy of $\sin^22\theta_{23}$ determination {\it does not} lead 
to the precise determination of $\theta_{23}$. 
It is due to the doubly bad luck we suffer, one by the large Jacobian at the 
nearly maximal mixing, and the other by the octant degeneracy of 
$\theta_{23}$, as discussed in detail in \cite{MSS}. 

On the other hand, the situation for $\theta_{12}$ is comfortably 
good as we have seen above. The SADO type experiment is feasible 
with modest requirement of 4\% for the experimental systematic error.
Then, the last message to the experimentalists; 
Why don't you attempt to carry out such experiment?

\vskip 0.2cm

This work was supported in part by the Grant-in-Aid for Scientific Research, 
No. 16340078, Japan Society for the Promotion of Science, 
and by 
Funda{\c c}{\~a}o de Amparo {\`a} Pesquisa do Estado de S{\~a}o 
Paulo (FAPESP) and Conselho Nacional de  Ci{\^e}ncia e Tecnologia (CNPq).


\begin{thebibliography}{99}


\bibitem {SKatm_new}
Y.~Ashie {\it et al.}  [Super-Kamiokande Collaboration],
%``Evidence for an oscillatory signature in atmospheric neutrino oscillation,''
Phys.\ Rev.\ Lett.\  {\bf 93} (2004) 101801.
%[arXiv:hep-ex/0404034].
%%CITATION = HEP-EX 0404034;%%


\bibitem{K2K_new}
T.~Nakaya, Talk at 
XXIst International Conference on Neutrino Physics and Astrophysics,  
June 14-19, 2004 Paris, France.


\bibitem {KamLAND_new} 
T.~Araki {\it et al.}  [KamLAND Collaboration],
%``Measurement of neutrino oscillation with KamLAND: Evidence of spectral
%distortion,''
arXiv:hep-ex/0406035.
%%CITATION = HEP-EX 0406035;%%


\bibitem{QLC} 
M. Raidal, Phys.\ Rev.\ Lett.\  {\bf 93}  (2004) 161801;
%[arXiv:hep-ph/0404046];  \\
%%CITATION = HEP-PH 0404046;%%
H.~Minakata and A.~Yu Smirnov, 
Phys.\ Rev.\ D {\bf 70}  (2004) 073009.
%[arXiv:hep-ph/0405088].
%%CITATION = HEP-PH 0405088;%%


\bibitem{PDG}
S.~Eidelman {\it et al.}  [Particle Data Group Collaboration],
%``Review of particle physics,''
Phys.\ Lett.\ B {\bf 592} 1 (2004) 1. 
%%CITATION = PHLTA,B592,1;%%

\bibitem{global}
M.~Maltoni, T.~Schwetz, M.~A.~Tortola, and J.~W.~F.Valle,
New J. Phys.  {\bf 6}  (2004) 122.
%[arXiv:hep-ph/0405172]. 
%%CITATION = HEP-PH 0405172;%%


\bibitem{bahcall-pena} 
J.~N.~Bahcall and C.~Pe{\~n}a-Garay, 
JHEP {\bf 0311} (2003) 004.
%[arXiv:hep-ph/0305159]. 
%%CITATION = HEP-PH 0305159;%%


\bibitem{MNTZ}
H.~Minakata, H.~Nunokawa, W.~J.~C.~Teves and R.~Zukanovich Funchal, 
Phys.\ Rev.\ D  {\bf 71}  (2004) 013005. 
%[arXiv:hep-ph/0407326]. 
%%CITATION = HEP-PH 0407326;%%


\bibitem {MSYIS} 
H.~Minakata, H.~Sugiyama, O.~Yasuda, K.~Inoue, and F.~Suekane, 
Phys.\ Rev.\ D {\bf 68}  (2003) 033017.
%[arXiv:hep-ph/0211111].
%%CITATION = HEP-PH 0211111;%%


\bibitem{reactor_white}
K.~Anderson  {\it et al.},
%White Paper Report on Using Nuclear Reactors to Search for a Value of $\theta_{13}$, 
arXiv:hep-ex/0402041. 
%%CITATION = HEP-EX 0402041;%% 


\bibitem {JPARC}
Y.~Itow {\it et al.}, arXiv:hep-ex/0106019. 
%%CITATION = HEP-EX 0106019;%%
%For an updated version, see: http://neutrino.kek.jp/jhfnu/loi/loi.v2.030528.pdf


\bibitem {MSS}
H.~Minakata, M.~Sonoyama and H.~Sugiyama, 
Phys.\ Rev.\ D {\bf 70}  (2004)  113012.
%[arXiv:hep-ph/0406073]. 
%%CITATION = HEP-PH 0406073;%%


\bibitem{BCGP}
A.~Bandyopadhyay {\it et al.},
%S.~Choubey and S.~Goswami,   
Phys.\ Rev.\ D {\bf 67}  (2003) 113011; 
%[arXiv:hep-ph/0302243]; 
%%CITATION = HEP-PH 0302243;%%
A.~Bandyopadhyay, S.~Choubey, S.~Goswami and S.~T.~Petcov,
%``High precision measurements of Theta(solar) in solar and reactor neutrino
%experiments,''
arXiv:hep-ph/0410283.
%%CITATION = HEP-PH 0410283;%%


\bibitem{KamLAND}
K.~Eguchi {\it et al.} [KamLAND Collaboration],
Phys.\ Rev.\ Lett.\  {\bf 90}  (2003) 021802.
%[arXiv:hep-ex/0212021].
%%CITATION = HEP-EX 0212021;%%


\bibitem{lowE}
M.~Nakahata, 
Talk at the 5th Workshop on 
``Neutrino Oscillations and their Origin'' (NOON2004), 
February 11-15, 2004, Odaiba, Tokyo, Japan.
http://www-sk.icrr.u-tokyo.ac.jp/noon2004/


\bibitem{borexino}
G.~Alimonti {\it et al.}  [Borexino Collaboration],
%``Science and technology of Borexino: A real time detector for low  energy
%solar neutrinos,''
Astropart.\ Phys.\  {\bf 16} (2002) 205.
%[arXiv:hep-ex/0012030].
%%CITATION = HEP-EX 0012030;%%


\end{thebibliography}
\end{document}